\documentstyle[aps,prl,floats,epsfig,twocolumn,amssymb]{revtex}

\begin{document}
\draft
\wideabs{
\title{Hall effect and conduction anisotropy in the organic conductor (TMTSF)$_2$PF$%
_6$}
\author{G. Mih\'{a}ly$^{1,2}$, I. K\'{e}zsm\'{a}rki$^{1,2}$, F. Z\'{a}mborszky$^{2}$
and L. Forr\'{o}$^{1}$}
\address{$^{1}$IGA, Ecole Politechnique Federale de Lausanne, CH-1015 Lausanne,\\
Switzerland\\
$^2$Department of Physics, Technical University of Budapest, H-1111\\
Budapest, Hungary \\
}
\date{Received 28 July 1999}
\author{}
\maketitle

\begin{abstract}
Long missing basic experiments in the normal phase of the anisotropic
electron system of (TMTSF)$_{2}$PF$_{6}$ were performed. Both the Hall
effect and the $ab^{\prime }$-plane conduction anisotropy are directly
addressing the unconventional electrical properties of this Bechgaard salt.
We found that the dramatic reduction of the carrier density deduced from
recent optical data is not reflected in an enhanced Hall-resistance. The
pressure- and temperature dependence of the $b^{\prime }$-direction
resitivity reveal isotropic relaxation time and do not require explanations
beyond the Fermi liquid theory. Our results allow a coherent-diffusive
transition in the interchain carrier propagation, however the possible
crossover to Luttinger liquid behavior is placed to an energy scale above
room temperature.
\end{abstract}

}


Recent extensive experimental investigations of the low-dimensional organic
conductor (TMTSF)$_{2}$PF$_{6}$ have revealed exciting electronic properties
in its metallic phase. The study of the optical conductivity over a wide
spectral range led to the puzzling result that the d.c. conduction along the
molecular chains ($a$-direction) is due to only about 1\% of the carriers,
and the dominant spectral weight is in a high frequency mode above a
correlation gap, $E_{gap}$ \cite{Vescoli,Dressel}. At low temperatures the
system is 2-dimensional (2d) as evidenced by the observation of a plasma
edge along the second most conducting ($b^{\prime }$) direction \cite
{Dressel}. The spin-density wave (SDW) phase transition at $T_{SDW}=12$ K is
the manifestation of a Fermi-surface anomaly and a broad variety of related
phenomena \cite{FL,Zambo1} is well understood in terms of the imperfect
nesting model of a 2d Fermi gas\cite{Montambaux}. However, with increasing
temperature a 2d $\longrightarrow $1d dimensionality crossover was
suggested, which results in decoupled chains exhibiting Luttinger-liquid
features \cite{Bourbonnais,Jerome}. The non-Fermi liquid behavior is
supported by NMR \cite{NMR}, magnetic susceptibility \cite{susceptibility}\
and photoemission \cite{photoemission} results. The power-low asymptotic
dependence of the high frequency optical mode has also been associated to
Luttinger-exponents \cite{Vescoli,Jerome}. The crossover temperature from a
low temperature Fermi-liquid to a high temperature Luttinger-liquid behavior
was identified by the resistivity peak measured along the least conducting ($%
c^{\ast }$) direction \cite{Jerome,Moser}.

The low temperature Fermi liquid state is generally described by a highly
anisotropic band structure with transfer integrals in the order of $%
t_{a}:t_{b}:t_{c}\approx 3000:300:10$ K. The crossover to a non-Fermi liquid
state has been related to various possible sources of electronic
confinement; \ i., to thermal energy exceeding the transverse coupling, $%
k_{B}T>$ $t_{b}$\ \cite{Bourbonnais}, ii., to the decrease of the transverse
mean free path below the separation of the chains, $l_{b}<b$, which leads to
the loss of coherence for the interchain transport \cite{Clarke}, and
finally iii., to a correlation gap exceeding the transfer integral
perpendicular to the chains, $E_{gap}>t_{b}$ \cite{Vescoli,Suzumura}. It has
also been suggested that in these relations a renormalized transfer integral
is relevant, $t_{b}^{eff}$, which may be substantially smaller than $t_{b}$%
\cite{Giamarchi}. The nature of the low temperature Fermi-liquid phase is
also controversial. Here we refer to the reduced number of the carriers
participating in d.c. transport, to the presence of a gapped high frequency
mode, and also to the fact that the unusually long relaxation time of $\
\tau \approx 2\ast 10^{-11}$ s \cite{Vescoli} corresponds to an anomalously
long mean free path along the chains, $l_{a}(20K)>10$ $\mu m$.

In order to get insight into the above exotic electrical features of the
metallic phase in (TMTSF)$_{2}$PF$_{6}$ two very basic experiments were
carried out for the first time (Hall-effect and $ab^{\prime }$-plane
anisotropy). The number of carriers participating in the electrical
transport is deduced from the Hall measurements performed in the temperature
range of $5-270$ K. The temperature and pressure dependence of the
transverse resistivity along the second most conducting direction, $\rho
_{b} $, was also determined. The results do not reveal evidence for
dimensionality crossover up to room temperature and the temperature
independent $ab^{\prime }$-plane anisotropy suggests isotropic in-plane
relaxation time.

The Hall effect was measured in a $B\parallel a$ configuration (magnetic
field parallel to the most conducting direction). The current ($I$) was
applied along the $c^{\ast }$-direction, while the Hall-voltage was measured
along the $b^{\prime }$-direction. Though the experimental realization of
such a configuration is hard, it has several advantages. First, in any other
configuration the Hall-sensing contacts are placed on well conducting
surfaces and thus they are short-circuited by the current injecting pads
which cover the ends of the crystal. This can be avoided by
surface-injecting (8-point configuration), but then the current may well be
inhomogeneous. Second, as in our case $I$ $\parallel c^{\ast }$ the
homogeneous current distribution is ensured by the almost equipotential $%
ab^{\prime }$-planes. Finally, the application of the magnetic field does
not induce Lorentz force pointing to the $c^{\ast }$-direction, along which
the developement of a Hall field may be influenced by the anions separating
the\ chains.

\begin{figure}[t]
\noindent
\centerline{\includegraphics[width=0.55\columnwidth]{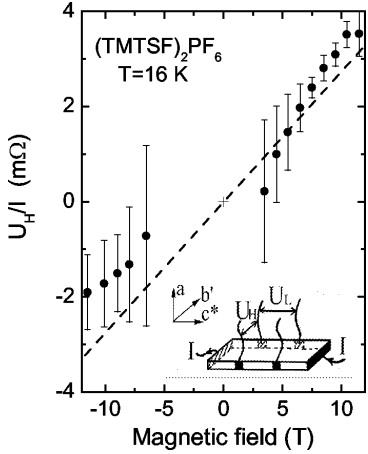}}
\vspace{1truemm}
\caption{Magnetic field dependence of the Hall voltage in the metallic phase
of (TMTSF)$_{2}$PF$_{6}$ (for details see text).}
\label{fig1}
\end{figure} 

The results displayed below were obtained on a narrow slice cut from a thick
crystal perpendicular to the chain axis. The resulting small piece was 700 $%
\mu m$ of length ($c^{\ast }$-direction), 650 $\mu m$ of width ($b$%
-direction) and had a thickness $d=$ 25 $\mu m$ ($a$-direction). The
contacts were made by evaporating gold pads on the $b^{\prime }c^{\ast }$ \
surface of the crystal, and then 6 $\mu m$ gold wires were attached by
silver paint. In the 6 contact arrangement applied, the 2 pairs of side
contacts allowed the simultaneous measurement of the Hall-voltage and of the 
$c^{\ast }$-direction resistance. Measurements on three other crystals gave
results consistent with those shown later, however due to their larger
thickness they had worse signal/noise ratio.

In order to avoid any systematic error in the Hall voltage (especially the
mixing of magnetoresistive components) both the current- and the
field-direction were rotated. The inversion of the magnetic field with
respect to the sample was achieved by rotating the crystal by a stepping
motor. Most experiments were performed in persistent mode at $B=12$ T, thus
with a fast 180$^{\circ }$ rotation of the sample we were able to produce a
magnetic field change of $\Delta B=24$ T within 1 second. This was followed
by a waiting time of $t_{th}=2$ s, to let the sample thermalized. Then the
Hall resistance, $R_{H}$, was measured with inverting currents and applying
digital averaging over $t_{av}=1$ s. This rotation-detection procedure was
repeated continuously, and the signal/noise ratio was further improved by
taking the average of the data obtained in about 10-50 cycles (over 1-5
minutes).

The above method was tested in several steps. The thermalization time was
chosen long enough to avoid temperature reading error due to the heating of
the sample by the Eddy currents (the good thermalization is shown for
example by the correct transition temperature determined from the Hall
data). Thermal drift or thermal gradient related to the rotations have not
influenced the data either. It was tested by changing the order both in the
sample rotation sequence and in the current inversion sequence, and the same
result were obtained in any of the four combinations. The reproducibility of
the data during the cooling and heating cycle was also confirmed. Finally we
excluded the possibility of any spurious systematic mechanical error (e.g.
the gears applied in the sample holder may have play of about 2$^{\circ }$).
For this aim we changed the magnetic field of the solenoid from $B=12$ T to $%
-12$ T \ step by step and measured the Hall-signal, as described above. This
supplied a double check for the inversion of the measured voltage with
inverting field. Furthermore, this experiment revealed that the Hall signal
is linear up to $12$ T, as shown in Fig. 1. (The slight curvature in the
Hall voltage can be attributed to the temperature drift occurred during the
more than 1 hour measuring time.)

\begin{figure}[h]
\noindent
\centerline{\includegraphics[width=0.9\columnwidth]{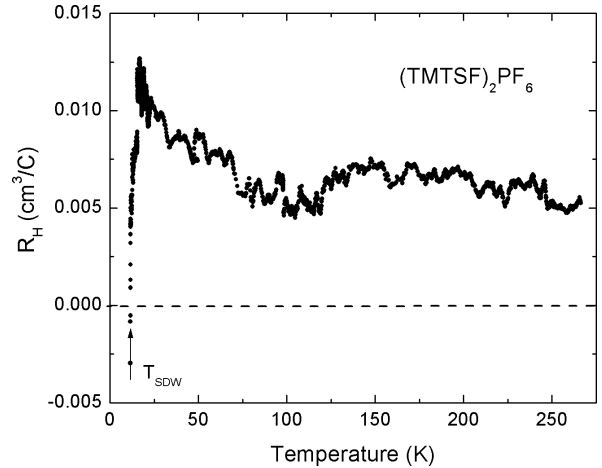}}
\vspace{1truemm}
\caption{Temperature dependence of the Hall resistance in the normal phase
of (TMTSF)$_{2}$PF$_{6}$. The carrier density of 1 hole/unit cell
corresponds to $R_{H}=4\times 10^{-3}$ cm$^{3}$/C. The error in the
magnitude of the Hall resistance is determined by the accuracy of the
thickness measurement ($\pm $30\%)}
\label{fig2}
\end{figure}

Figures 2 and 3 show the temperature dependence of the Hall-resistance of
(TMTSF)$_{2}$PF$_{6}$ in the normal and in the SDW phase, respectively. In
the metallic state the Hall signal is small, in accordance with the early
observation of Jacobsen et al. who determined only an upper limit for it ($%
\left| R_{H}\right| \lessapprox 10^{-2}$ cm$^{3}$/ C in $B\perp a$
configurations)\cite{Jacobsen}. We found that the Hall resistance is
temperature independent, except in the vicinity of the phase transition. At
the transition temperature it changes sign, then in the SDW phase $\left|
R_{H}\right| $ rapidly increases 3 orders in magnitude down to $T=5$ K. As
expected for a semiconductor, the Hall resistance is activated. The
activation energy is $\Delta =23$K, in accordance with previous results
obtained in $B\parallel b^{\prime }$\ configuration \cite{Chaikin}. Note,
however, that the magnitude of $\left| R_{H}\right| $ measured in the SDW
phase when $B\parallel a$ is significantly smaller than that found either
for the positive Hall coefficient in the $B\parallel b^{\prime }$\ or for
the negative Hall resistance in the $B\parallel c^{\ast }$\ alignments \cite
{Chaikin,Forro}.

\begin{figure}[h]
\noindent
\centerline{\includegraphics[width=0.75\columnwidth]{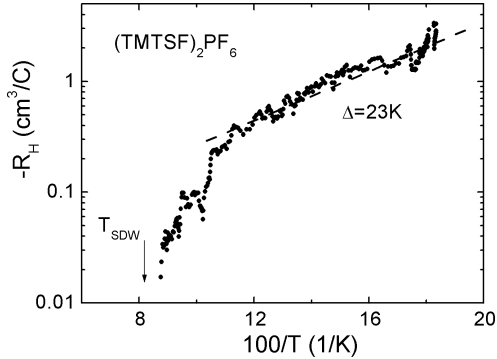}}
\vspace{1truemm}
\caption{Arrhenius plot of the Hall resistance in the SDW phase of (TMTSF)$%
_{2}$PF$_{6}$.}
\label{fig3}
\end{figure}

The resistivity, $\rho (T)$, along the different directions were measured in
various four probe arrangements, at samples cut from a thick crystal. In
case of $\rho _{b}$ four gold strips were evaporated on the $b^{\prime
}c^{\ast }$ surface of the crystal, while for $\rho _{c}$ two pairs of
contacts were placed on the opposite $ab^{\prime }$-surfaces. Typical size
of these samples was $0.4\times 0.4\times 0.1$ mm$^{3}$ ($c\times b\times a$%
). The $ab^{\prime }$-plane anisotropy was determined by Montgomery method 
\cite{Montgomery}, as well. In this case the contacts were put on four
corners of the $ab^{\prime }$ side of a long crystal, by placing the gold
pads on the opposite $ac^{\ast }$ surfaces. We verified that $\rho
_{a}(T)/\rho _{b}(T)$ calculated from two independent longitudinal
measurements agrees with the temperature dependence of the anisotropy
determined by the Montgomery method on several crystals; and vica-versa, $%
\rho _{a}(T)$ or $\rho _{b}(T)$ calculated from the Montgomery measurements
agree with the results of the direct longitudinal measurements.

Figure 4 shows the temperature dependence of the resistivity measured along
the different directions. The $a$- and $c^{\ast }$-direction data agree well
with those published in the literature by various groups. The $b^{\prime }$%
-direction data, however, differ from the single available result published
almost 20 years ago \cite{Jacobsen}. As plotted in Fig. 4, $\rho _{a}$ and $%
\rho _{b}$ exhibit the same temperature profile, the curves scale together.
In the normal phase the anisotropy is temperature independent within $20\%$,
its magnitude is $\rho _{b}/\rho _{a}=110\pm 30\%$. (In small specimens the
uncertainty of the contact positions introduces a large scatter in the
magnitude of the anisotropy; the above value is the average of 6 independent
measurements.)

\begin{figure}[t]
\noindent
\centerline{\includegraphics[width=0.85\columnwidth]{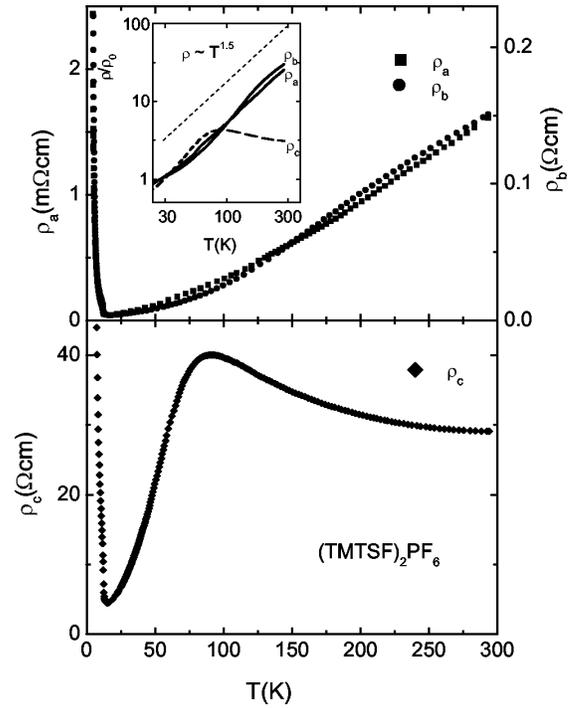}}
\vspace{1truemm}
\caption{ Temperature dependence of the resistivity measured along the
various crystallographic directions. The inset shows the curves on
logarithmic scales (normalized at $T=30$K), the dashed line corresponds to $%
\protect\rho \propto T^{1.5}$.}
\label{fig4}
\end{figure}

The pressure dependence of the resistivity along the $a$- and $b^{\prime }$%
-directions was also determined. The samples were inserted into a
self-clamping CuBe cell with kerosene as pressure medium. The results of a
Montgomery experiment are shown in Fig. 5 together with the direct $a$-axis
data. The pressure induced variation is the same for both orientations.

\begin{figure}[b]
\noindent
\centerline{\includegraphics[width=0.7\columnwidth]{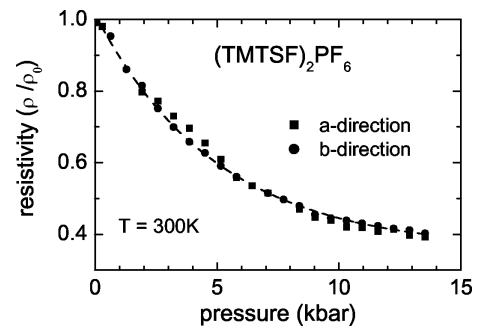}}
\vspace{1truemm}
\caption{Pressure dependence of the resistivity measured along the $a$- and
the $b^{\prime }$-directions.}
\label{fig5}
\end{figure}

In the discussion it will be outlined that all the above observations are
consistent with the Fermi-liquid description\cite{Kriza}. Starting with the
resistivity data, in contrast to the expectation of the Luttinger liquid
picture \cite{Moser,Jerome}, the temperature dependence along the $a$- and $%
b^{\prime }$-directions is similar from $T=30$ to $300$ K (Fig. 4). In a
good approximation $\rho _{a}(T)\propto \rho _{b}(T)\propto T^{\alpha }$
with \ $\alpha \approx 1.5$. Though the functional form may change if the
isobaric temperature dependence is transformed to constant volume data \cite
{Moser,Cooper0} such a transformation modifies $\rho _{a}$ and $\rho _{b}$
the same way, since they obey identical pressure dependences (Fig. 5).

The proportionality, $\rho _{a}(T)\propto \rho _{b}(T)$\ suggests a simple
anisotropic band structure with isotropic relaxation time, $\tau (T)$. In a
tight binding model a quarter filled band has a conduction anisotropy of: $%
\rho _{b}/\rho _{a}=(at_{a}^{2})/(bt_{b}{}^{2})$. This relation is certainly
valid at low temperature where the mean free path along the $b$-direction
exceeds the lattice constant: $l_{b}(20K)$ $\approx 20$ \AA \thinspace\ for $%
\tau \approx 4\ast 10^{-13}$ s \cite{Jacobsen}, $\rho _{b}/\rho _{a}\approx
100$ and Fermi velocity, $v_{F}\approx 4\ast 10^{-7}$cm/s \cite{Zambo1}.
With increasing temperature $l_{b}(T)$ decreases below the distance between
the chains ($7.7$ \AA ) at $T_{X}\approx 50$ K and above this temperature
the interchain carrier propagation becomes diffusive. For such an incoherent
interchain motion the perpendicular hopping probability is given by $\tau
_{b}^{-1}=\tau (t_{b}/\hslash )^{2}$ \cite{Weger}, ie. it is determined by
the lifetime along the chain direction. As a consequence, even a diffusive $%
b^{\prime }$-direction transport follows the temperature dependence of $\tau
(T)$, moreover the magnitude of the anisotropy is the same as in case of the
coherent $b^{\prime }$-direction transport (within a factor close to unity) 
\cite{Weger}.

A coherent-diffusive crossover has not been observed along the $c^{\ast }$%
-direction either, where the mean free path is much smaller \cite{Cooper1}.
The resistivity anomaly observed in $\rho _{c}(T)$ (Fig. 4a) can easily be
related to the fact that along the $c^{\ast }$-direction the chains are
separated by the PF$_{6}$ anions thus the transport may rather be
characteristic of the hopping process through the anions than of the nature
of an ideal anisotropic electron system \cite{Cooper1,Zambo2}. Note also
that around $60$ K the proton relaxation data indicate structural
rearrangement of the PF$_{6}$ ions \cite{Scott} thus the conclusions drawn
from $c^{\ast }$-direction transport \cite{Moser,Jerome} should be taken
cautiously.

The Hall effect is a quite general measure of the carrier concentration in
metals. The relation $R_{H}=1/ne$ is independent of the scattering
mechanisms (for isotropic relaxation time, as it is in our case) and remains
valid even for a strongly anisotropic band structure\cite{Cooper2}. As shown
in Fig. 2, the Hall coefficient observed in the normal phase of (TMTSF)$_{2}$%
PF$_{6}$ is close to the value $R_{H}=4\times 10^{-3}$ cm$^{3}$/C
corresponding to a carrier concentration of one hole/unit cell ( $%
n=1.4\times 10^{21}$ cm$^{-3})$. This is to be contrasted to the proposed
factor of 100 reduction of the concentration of the carriers participating
in the d.c. transport \cite{Vescoli,Dressel}. Such a reduction should lead
to a factor of hundred enhancement in $R_{H}$, which obviously has not been
observed.

The enhancement of the Hall-signal in the vicinity of the phase transition
can well be attributed to the opening of a pseudogap due to precursor
fluctuations. In the semiconducting phase our results reflect the
exponential freezing out of the carriers and the activation energy agrees
well with that obtained from the resistivity data. While previous Hall data
of $B\parallel b^{\prime }$\ or $B\parallel c^{\ast }$\ configurations \cite
{Chaikin,Forro} led anomalously large Hall-mobilities in the SDW phase
(explained by introducing in-chain effective mass as small as 1/200 m$_{e}$ 
\cite{Chaikin}) our data is consistent with a mean free path of about $%
\lambda _{a}(5K)\approx 50$ \AA\ for $\ m\approx m_{e}$.

In conclusion we performed basic transport experiments in the normal phase
of (TMTSF)$_{2}$PF$_{6}$. The results do not require explanations beyond the
Fermi liquid description. The Hall effect corresponds to a carrier density
of 1 hole/unit cell and the huge enhancement, expected from the optical data 
\cite{Vescoli,Dressel}, was not found. In contrast to previous suggestions
based on the $c$-direction transport \cite{Moser} $\rho _{b}(T)$ does not
confirm the Luttinger liquid picture. With increasing temperature a
coherent-diffusive transition occurs along the $b^{\prime }$-direction at $%
T_{X}\approx 50$ K, however this is a smooth crossover and it does not show
up in the anisotropy. The incoherent interchain transport still allows
strong coupling between chains and for the conduction the conventional Fermi
liquid description \cite{Weger} remains valid. Our results place the
possible appearance of the 1d Luttinger liquid features in the dc transport
to a higher energy scale (above room temperature); $k_{B}T>$ $t_{b}$, the
bare transfer integral or $k_{B}T>$ $E_{gap}$, the correlation gap observed
in optical data.

We thank B. Alavi for sample preparation and G. Gruner, S. Brown, L.
Degiorgi and M. Dressel for useful discussions. This work was supported by
the Swiss National Foundation for Scientific Research and by Hungarian
Research Funds OTKA T015552, FKFP 0355-B10.

$Note$ $added$. -- After submission of this article we received a preprint
reporting similar Hall results in the normal phase of (TMTSF)$_{2}$PF$_{6}$ 
\cite{Moser2}.

\bigskip


\end{document}